\documentclass[aps,prl,twocolumn,showpacs]{revtex4}
\usepackage{amsmath}
\usepackage{graphicx,epsfig,psfrag}
\usepackage{amssymb}

\renewcommand{\vec}[1]{{\bf #1}}
\begin{document}
\title{Kondo proximity effect: How does a metal penetrate into a Mott insulator?}
%\title{How a metal penetrates into a Mott insulator}
% How insulating is a Mott insulator?
% 
 
\author{R. W. Helmes${}^1$}
\author{T. A. Costi${}^2$}
\author{A. Rosch${}^1$}
\affiliation{\hspace*{-1.9cm}${}^1$Institute for Theoretical Physics, University of Cologne, 50937
Cologne, Germany\hspace*{-1.9cm} \\
${}^2$Institute of Solid State Research, Research Centre J\"ulich, 52425 J\"ulich, Germany}
\begin{abstract}
  We consider a heterostructure of a metal and a paramagnetic Mott
  insulator using an adaptation of dynamical mean field theory to
  describe inhomogeneous systems. The metal can penetrate into the
  insulator via the Kondo effect. We investigate the scaling
  properties of the metal-insulator interface close to the critical
  point of the Mott insulator. At criticality, the quasiparticle
  weight decays as $1/x^2$ with distance $x$ from the metal within our
  mean field theory. Our numerical results (using the numerical
  renormalization group as an impurity solver) show that the prefactor
  of this power law is extremely small.
\end{abstract}

\vspace{-0.8cm}

%73.20.-r Electron states at surfaces and interfaces
%71.27.+a Strongly correlated electron systems; heavy fermions
%71.30.+h 	Metal-insulator transitions and other electronic transitions 

\date{\today}
\maketitle

{\em Introduction:} In the last few years, an enormous amount of
interest has arisen in heterostructures fabricated out of strongly
correlated materials.  Driven by the prospect of new effects and
devices based on correlated electron compounds, a wide range of
systems has been studied experimentally and theoretically. For
example, the interface of two Mott insulators can show metallic
behavior \cite{Ohtomo2002,Mannhart2006,OkamotoMillis,Lee2006,Kancharla2006}
or can even become superconducting \cite{Mannhart2007}. The conducting
layers can, for example, arise from interface charges induced by the
Coulomb interactions. As shown by Thiel {\it et al.}
\cite{Mannhart2006} such interfaces can be manipulated by gate
voltages thereby opening the prospect for interesting novel devices.

The physics of such inhomogeneous systems can also play a role for the
properties of bulk materials where competing phases lead to the
formation of domain walls. For example, it has been argued in
Ref.~\cite{papa08} that the conductivity close to the endpoint of the
first-order Mott transition in certain organic salts of the
$\kappa$-ET family is dominated by domain wall effects. Also for cold
atoms, the trapping potential naturally makes the experimental systems
inhomogeneous which often leads to coexisting phases and corresponding
phase boundaries \cite{Helmes08}.

In this paper we will investigate the interface of a metal and a Mott
insulator. How does a metal penetrate into a Mott insulator?  The main
difference between a Mott insulator and an ordinary band insulator is
the presence of magnetic degrees of freedom arising from the localized
spins. While the large charge gap, of the order of the local Coulomb
repulsion $U$, prohibits tunneling of electrons into a Mott insulator,
the resonant spin flip scattering opens a new channel for tunneling
via the well-known Kondo effect and allows metallic behavior to be
induced within the Mott insulator.  Due to this `Kondo proximity
effect' an insulating layer adjacent to the metal will also become
metallic. In this manner the metal `eats' itself layer by layer into
the Mott insulator if not stopped either by magnetism or thermal
fluctuations. Here, we study this physics within the simplest setup
consisting of a particle-hole symmetric Hubbard model (see below)
where the local interaction $U$ jumps across the interface from
$U=U_\text{left} \ll U_c$ to a value $U=U_\text{right}$ close to the
critical coupling $U_c$ of the Mott transition.  For such a model the
charge is always homogeneous and no complications due to
charge-reconstruction or charged surface layers arises. Furthermore,
we only consider a paramagnetic Mott insulator and comment on the role
of magnetism only in the conclusions.

For one-dimensional systems, powerful numerical methods like DMRG are
available to study inhomogeneous strongly interacting models
\cite{Oka2005}.  For three dimensional systems, however, further
approximations are necessary. Here, the method of choice to study the
Mott transition is the so-called dynamical mean field theory (DMFT)
\cite{MetznerVollhardt,GeorgesRevMod1996}. Within DMFT, the only
approximation is to neglect non-local contributions to the
self-energy. This approximation can be used both for homogeneous
\cite{MetznerVollhardt,GeorgesRevMod1996} and inhomogeneous
\cite{Dobrosavljevic1997,Potthoff,Freericks,Florens,OkamotoMillis} problems.  In
the case of a heterostructure, each layer is effectively mapped to a
single-impurity Anderson model. These are coupled by a
self-consistency condition, see below.

Potthoff and Nolting used this spatially resolved DMFT to study the
Mott transition at surfaces using the semi-infinite Hubbard model
\cite{Potthoff1999}. In this context, also the question was studied
how a metallic surface influences the insulating bulk. Spatially
resolved DMFT has also been applied
\cite{Freericks,OkamotoMillis,Lee2006} to
investigate heterostructures of Mott insulators and band insulators.

A main problem of DMFT is the need for a reliable and efficient method
to solve the effective impurity problem. Previous applications of DMFT
to inhomogeneous systems used impurity solvers based on exact
diagonalization of small systems \cite{Potthoff1999}, a linearized
version of DMFT close to the critical point \cite{Potthoff1999}, a
two-site approximation \cite{OkamotoMillis,Lee2006} or slave-boson
mean-field theory \cite{Dobrosavljevic1998}, implying severe further
approximations, or started from simpler models such as the
Falicov-Kimball model \cite{Freericks}. Only recently
\cite{Helmes08,Hofstetter08}, the numerical renormalization group
(NRG) method \cite{BullaPRB2001,Bulla2007} was implemented as an
impurity solver to study the Mott transition of trapped atoms in an
optical lattice. We will also use this approach here as the NRG
appears to be the only method presently available which can
quantitatively resolve quasiparticle weights as small as $10^{-3}$
which are needed to describe the physics close to the Mott transition.

After introducing the model and our method (DMFT for inhomogeneous
systems+NRG), we will first investigate the heterostructure at finite
$T$ and for $U_\text{right}=U_c$. We will then analyze the $T=0$
scaling properties of the interface region at $U_\text{right} \lesssim
U_c$ and $U_\text{right} \gtrsim U_c$ using both DMFT+NRG and a
Ginzburg-Landau type analysis.

{\em Model and Method:} To investigate the junction of the metal and
the Mott insulator, we will consider the half-filled Hubbard model
\begin{align}
  {\cal{H}} =- t \sum_{\langle ij \rangle,\sigma} c_{i\sigma}^{\dag}
  c_{j\sigma} + \sum_i U_i (n_{i\uparrow}-\frac{1}{2})
  (n_{i\downarrow} -\frac{1}{2})
\end{align}
on a three-dimensional cubic lattice with the half band width $D=6 t$.
While we will consider a uniform hopping $t$, we choose
$U_i=U_\text{left}=D$ for $x \le 0$ describing a metal with a sizable
quasiparticle weight $Z_{\rm metal}=0.62$. For sites with $x \ge 1$,
we use an interaction $U_i=U_\text{right} \sim U_c$ close to the the
critical value, $U_c \approx (2.79 \pm 0.01) D$ which separates the
metallic from the insulating phase in the bulk. Note that the system
is translationally invariant in the $yz$ directions as $U_i$ is
constant within each $yz$ layer.

The DMFT algorithm for this heterostructure is almost identical to the
standard one \cite{GeorgesRevMod1996}. As all sites within a single
$yz$ layer are equivalent, it is sufficient to solve only one
effective Anderson impurity problem (using NRG \cite{Bulla2007,NRG})
for each $yz$ layer to obtain an $x$ dependent self energy
$\Sigma_x(\omega)$. From this one obtains the lattice Greens function
\begin{eqnarray}\label{lat}
  \hat{G}^\text{lat}(\epsilon_{\vec{k}_\perp},\omega)=
  \left(\omega-\epsilon_{\vec{k}_\perp}-t_{xx'}-\delta_{xx'}\Sigma_x(\omega)\right)^{-1}
\end{eqnarray}
written as a matrix in the $x$ coordinates where $t_{xx'}=t$ for $x' =
x\pm 1$ and 0 otherwise and $\epsilon_{\vec{k}_\perp}$ is the
dispersion within each layer. From $\hat{G}^\text{lat}$ one determines
the {\em local} Greens function which is used \cite{GeorgesRevMod1996}
to derive a new effective Anderson impurity model with the Greens
function $G_\text{imp}(x)$ for each layer using the self-consistency
equation

\begin{eqnarray}\label{selfcon}
  G^{\text{imp}}_x(\omega)\stackrel{!}{=}\int d\epsilon_{\vec{k}_\perp} N_{2d}(\epsilon_{\vec{k}_\perp})
  \left. \hat{G}^\text{lat}(\epsilon_{\vec{k}_\perp},\omega)\right|_{xx} ,
\end{eqnarray}
where $N_{2d}(\epsilon)=\sum_{\vec{k}_\perp}
\delta(\epsilon-\epsilon_{\vec{k}_\perp})$ is the two-dimensional
density of states of the $yz$ layers. To avoid numerical difficulties
associated with the logarithmic divergence of 2d cubic density of
states, we use $N_{2d}(\epsilon)=1/(8 t)$ for $|\epsilon|<4 t$. This
does not affect any universal properties discussed below and leads
only to small changes in $U_c$ (and other numerical prefactors) of
about 10\%. From Eq. (\ref{lat}) it seems that one has to invert a
matrix for each value of $\omega$ and $\epsilon_{\vec{k}_\perp}$.
Fortunately, this computationally expensive step can be simplified by
diagonalizing $\hat{G}^\text{lat}(0,\omega)^{-1}$ using the orthogonal
matrix $\hat{O}(\omega)$, $\hat{O}^T(\omega)
\hat{G}^\text{lat}(0,\omega)^{-1} \hat{O}(\omega) = \hat{M}(\omega)$,
so that $\hat{G}^\text{lat}(\epsilon_{\vec{k}_\perp},\omega)=
\hat{O}(\omega)(\hat{M}(\omega)-\epsilon_{\vec{k}_\perp})^{-1}
\hat{O}^T(\omega)$. We use $20$ metallic layers with $U=D$ and $40$
layers with $U \sim U_c$ which is sufficiently large to avoid any
finite size effects.

\begin{figure}
\includegraphics[width=\linewidth,clip]{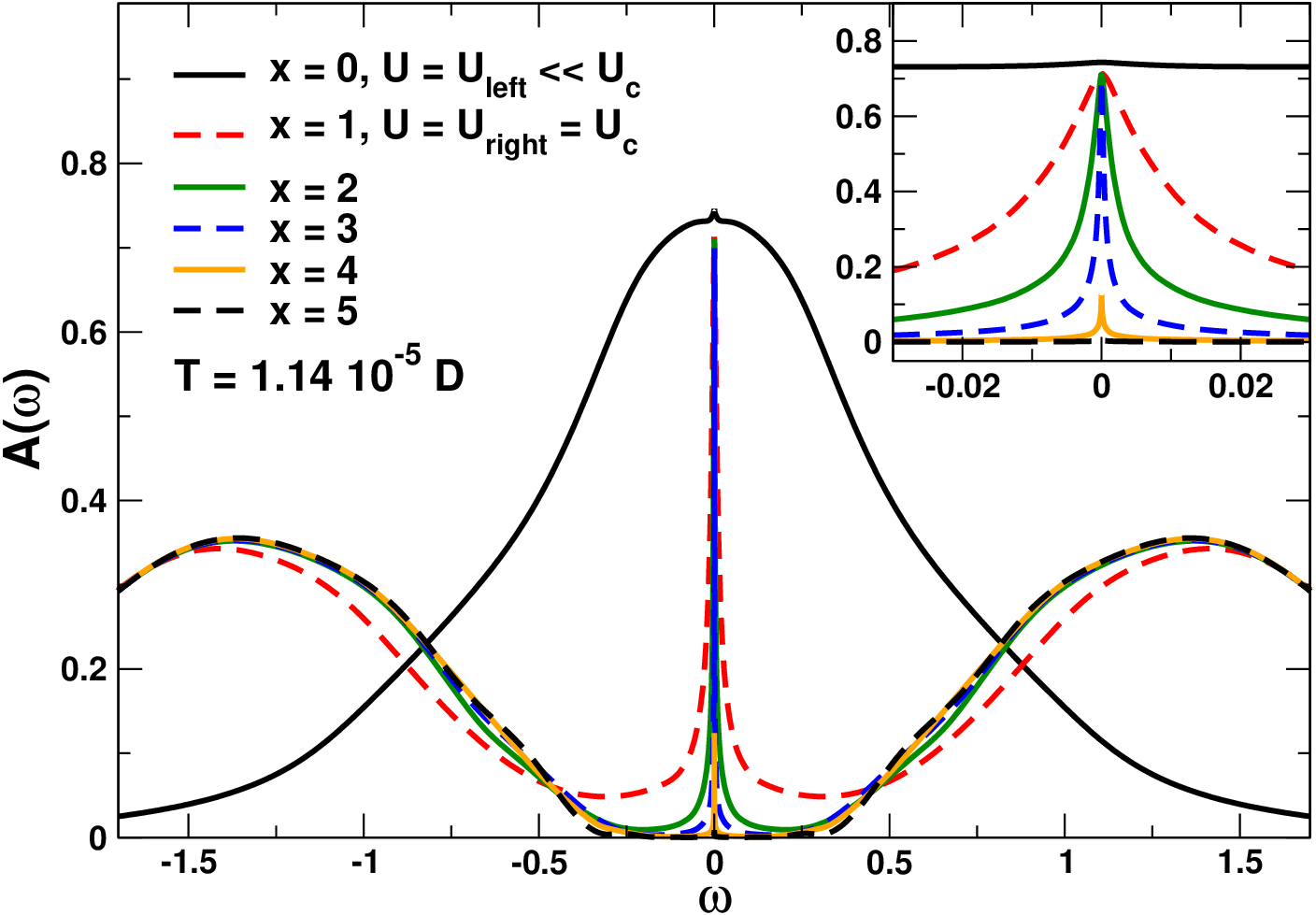}
\caption{{\em (Color online)} Layer dependence of local spectral
  function close to the interface $x=0$ for $U_\text{right}=U_c$ and
  $T= 1.14\,\times 10^{-5}\,D$.  Inset: $A(\omega)$ near $\omega=0$.
  \label{fig1}}
\end{figure} {\em Results:} Fig. \ref{fig1} shows the layer dependence
of the spectral function for $U_\text{right}=U_c$ for $x\ge 0$ and for
a low temperature $T= 1.14\,\times 10^{-5}\,D$.  The metallic side is
only weakly affected by the presence of the insulator as within our
paramagnetic, particle-hole symmetric model no Friedel oscillations
occur.
%(e.g., the hybridization with the sharp features for $x\ge 1$ leads
%to a tiny hump in the  $x=0$ curve)
All layers with $U=U_c$ show pronounced Hubbard bands. The width of
the sharp quasiparticle peak, which describes the penetration of the
metal into the quantum critical Mott state, decays rapidly. The
quasiparticle peak collapses completely (up to an exponentially small
feature) from the 5th layer on, when the Kondo temperature of the
corresponding impurity model becomes much smaller than $T$.

\begin{figure}
\includegraphics[width=\linewidth,clip]{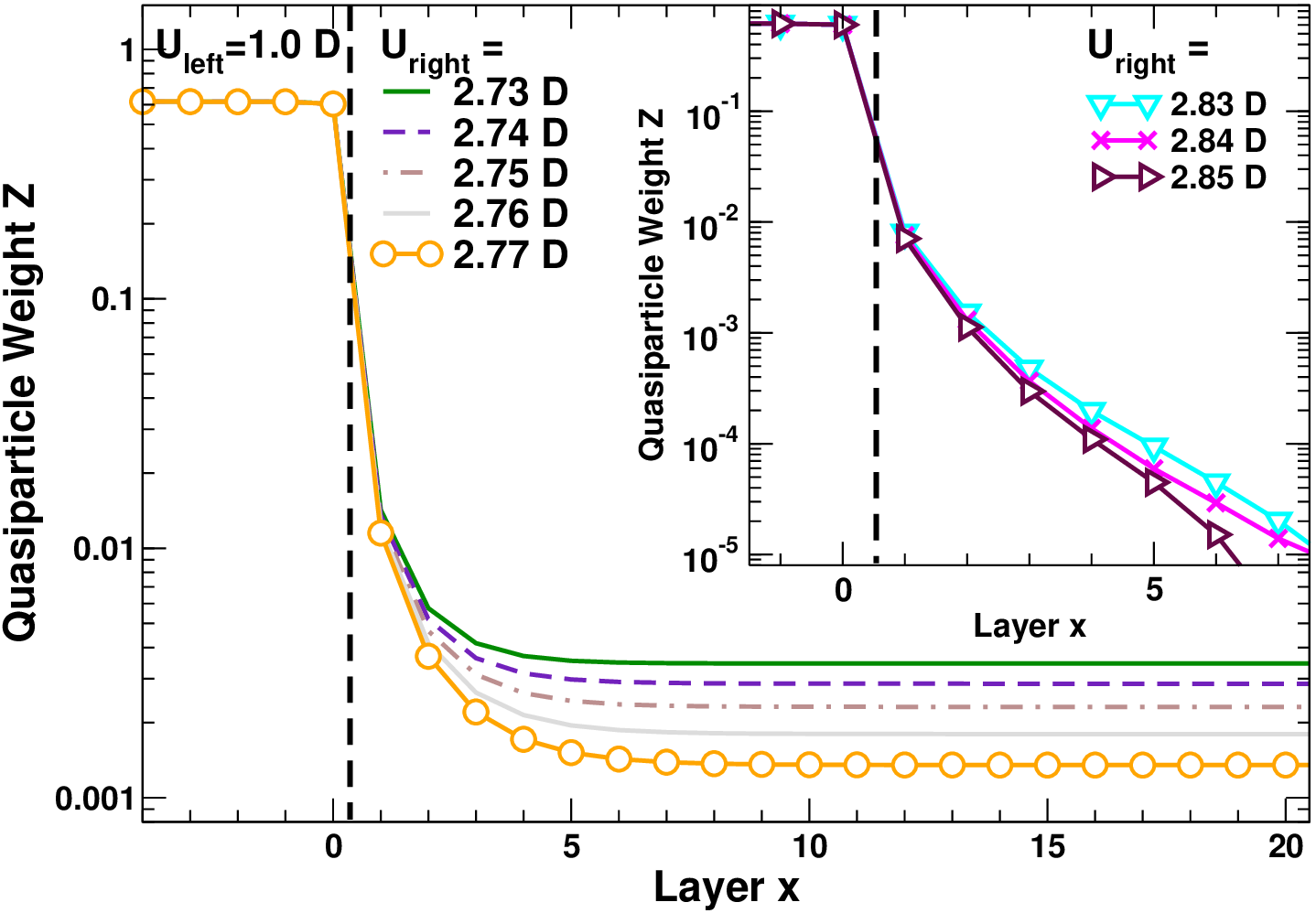}
\caption{{\em (Color online)} Quasiparticle weight $Z$ for an
  inhomogeneous layered system describing a `good' metal
  ($U=U_{\text{left}}=D$ for $x <1$) in contact with a `bad' metal
  ($U_\text{right}\lesssim U_c=(2.79 \pm 0.01) D$ for $x \ge 1$).
  Inset: Quasiparticle weight for a heterostructure of a good metal
  and a Mott insulator ($U_\text{right}\gtrsim U_c$). In both cases,
  $Z$ drops as $1/x^2$ for $x < \xi$, see Fig.~\ref{fig3}.  For
  extremely small quasiparticle weights $Z<10^{-3}$ the numerical
  results eventually become unreliable.\label{fig2}}
\end{figure}

\begin{figure}
\includegraphics[width=\linewidth,clip]{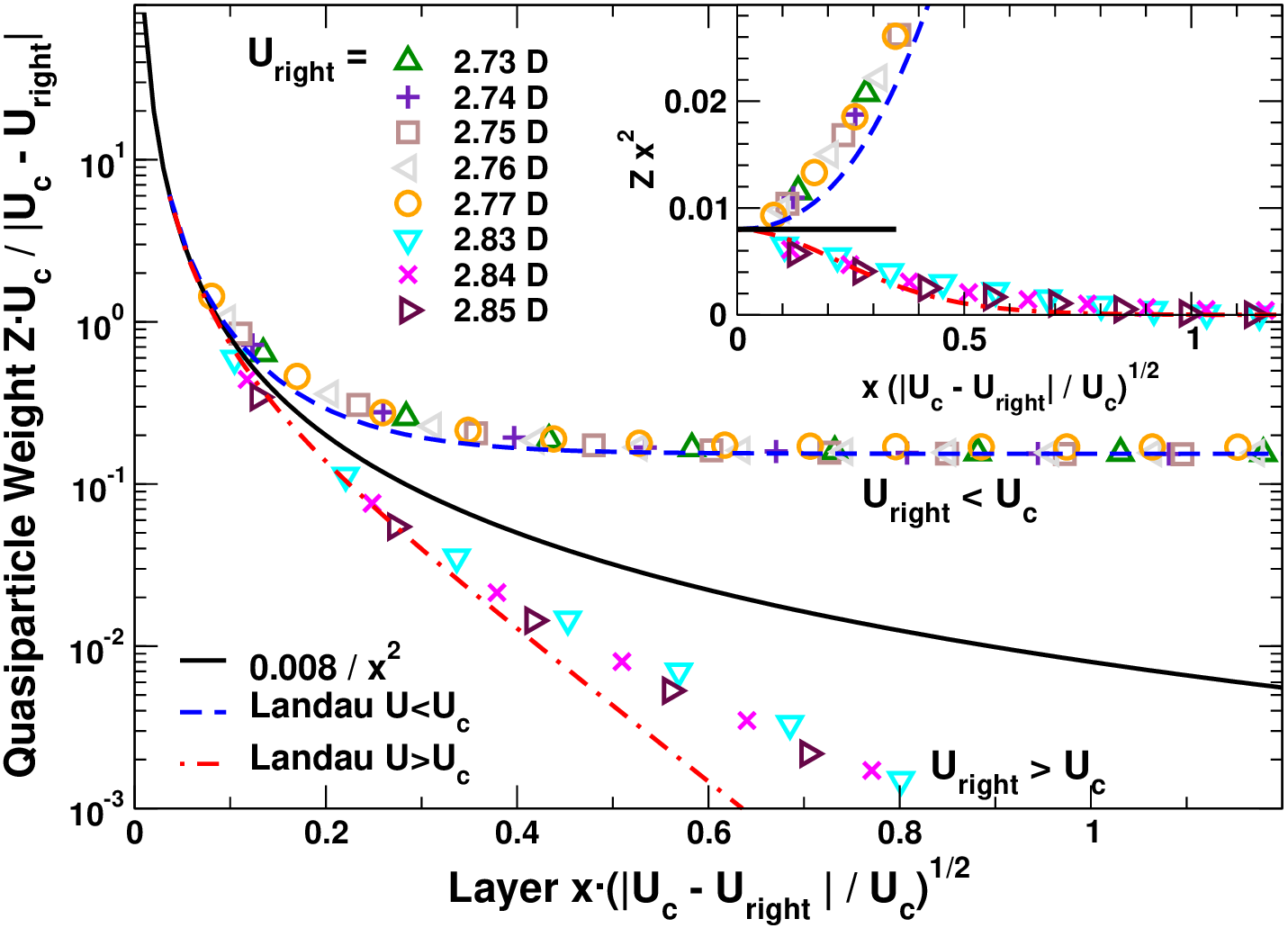}
\caption{{\em (Color online)} Scaling plot \cite{shift} of the $T=0$
  quasiparticle weight close to the quantum critical point both for
  $U_\text{right}<U_c$ (upper curves) and $U_\text{right}>U_c$ (lower
  curves) confirming $\nu=1/2$ in Eq.~\ref{scaling}.  In the
  quantum-critical regime ($x < \xi$), $Z \approx 0.008/x^2$ (solid
  line). For $U_\text{right}<U_c$, $Z$ saturates at a finite value
  proportional to $U_c-U_\text{right}$ for $x > \xi$ while it drops
  exponentially for $U_\text{right}>U_c$, see Eq.~\ref{qpIns}. Dashed,
  dot-dashed lines: scaling curves obtained from a Ginzburg-Landau
  type analysis using Eqs.~(\ref{zp}) and (\ref{z}). The results for
  $Z U_c/(U_\text{right}-U_c) < 0.05$ are not numerically reliable. 
  \label{fig3}}
\end{figure}

For a quantitative analysis of how the metal penetrates into the Mott
insulator we investigate a heterostructure consisting of a 'good
metal', $U_\text{left}=D$ and a 'bad metal', $U_\text{right} \lesssim
U_c$ at $T=0$. For $T=0$ the quasiparticle weight $Z_x$ of layer $x$
is well defined and can be obtained from $Z_x=(1-\partial_\omega
\text{Re} \Sigma_x(\omega))^{-1}$. Fig.~\ref{fig2} shows the
quasiparticle weight $Z$ as a function of the distance $x$ from the
interface. Upon increasing $U_\text{right}$ the quasiparticle weight
deep in the bad metal decreases linearly with $U_c-U_\text{right}$
\cite{GeorgesRevMod1996}. Close to the critical point one expects
scaling behavior and indeed we observe in Fig.~\ref{fig3}
\begin{eqnarray}\label{scaling}
  Z_x \approx \frac{0.008 \pm 0.002}{x^{1/\nu}} 
  f\!\left[x \left(\frac{U_c-U_\text{right}}{U_c}\right)^{\nu}\right]
\end{eqnarray}
with $\nu=1/2$ where $f[u]$ is an universal scaling function with
$f[0]=1$ and $f[u\to \infty]\approx (0.150 \pm 0.005) u^2$ for
$U_\text{right}\lesssim U_c$. The observation that DMFT is
characterized by the usual mean-field exponent $\nu=1/2$ and the
$1/x^2$ decay of the correlation function in the quantum critical
regime is one of the main results of this paper. Defining the
correlation length $\xi$ by $f[u]=2$, we obtain
\begin{equation}
  \xi \approx 0.3 \left(\frac{U_c}{U_c-U_\text{right}}\right)^{1/2}.
\end{equation}

In the Mott insulating phase at $T=0$, $U_\text{right} \gtrsim U_c$,
the quasiparticle weight drops rapidly, see inset of Fig.~\ref{fig2}.
In this regime, the numerical calculations become difficult and are
unreliable for $Z < 10^{-3}$.  Nevertheless, in the regime where the
numerical results are accurate, they confirm the scaling ansatz
Eq.~({\ref{scaling}}), as can be seen in detail in Fig.~\ref{fig3}: In
the quantum-critical regime, i.e. for $x<\xi$, the quasiparticle
weight decays as $1/x^2$ with the {\em same} prefactor as in
Eq.~(\ref{scaling}). For $x > \xi$, however, $Z$ drops exponentially
but remains always finite.

This picture is further corroborated by an analysis in the spirit of a
Ginzburg-Landau mean-field treatment as in
\cite{zhang1993,Bulla1999,Potthoff1999}. The basic idea is that close
to $U_c$, the physics is mainly determined by the quasiparticle peak
which can be characterized by a single number, the quasiparticle
weight $Z_x$. We therefore approximate in Eq. (\ref{lat})
$\Sigma_x(\omega)\approx \omega-\omega/Z_x$. The resulting local
spectral function $-\frac{1}{\pi}\operatorname{Im}
G^\text{lat}_{xx}(\omega)$ has a peak with a finite width. From this
peak one has to determine a single number describing the effective
impurity model and from this a new value of $Z_x$ using a
Ginzburg-Landau expansion around the critical point.
\begin{eqnarray}
  Z'_x&=&\frac{3}{22}(Z_{x-1}+\frac{16}{3} Z_x+Z_{x+1}) \label{zp}
\end{eqnarray}
\begin{eqnarray}
  Z_x&=& Z'_x-\alpha \frac{U-U_c}{U_c} Z'_x - \beta {Z'_x}^2 \label{z}
\end{eqnarray}
For the first step, Eq. (\ref{zp}), we have used the procedure
described by Bulla, Potthoff and Nolting \cite{Bulla1999,Potthoff1999}
with $\int N_{2d}(\epsilon) d\epsilon = 16 t^2/3$. For the homogeneous
system with $U<U_c$ one obtains $Z= \frac{\alpha}{\beta}
\frac{U_c-U}{U_c}$. Analyzing the asymptotic solutions of
Eqs.~(\ref{zp}) and (\ref{z}) for the interface with
$U_{\text{right}}=U_c$ we find $\lim_{x \to \infty} Z_x=9/(11 \beta
x^2)$. The model (\ref{zp},\ref{z}) reproduces the critical exponents
of our DMFT calculation (\ref{scaling}). Fitting the asymptotic
formulas to the NRG results we obtain $\alpha=15.7 \pm 5$ and
$\beta=102 \pm 30$, reflecting the small prefactor in
Eq.~(\ref{scaling}). While Potthoff and Nolting \cite{Potthoff1999}
have analyzed different critical exponents, their results are
qualitatively fully consistent with ours.

In Fig.~\ref{fig3}, we show that the Ginzburg-Landau analysis
reproduces the DMFT scaling curves quite well as the parameters in
(\ref{zp},\ref{z}) have been determined to fit the prefactor of the
$1/x^2$ law and the large $x$ limit for $U_\text{right}<U_c$. The
large deviations for $U_\text{right} > U_c$ arise in the
aforementioned regime where the numerical calculations are no longer
reliable.  There is also a smaller deviation for $U_\text{right} <
U_c$ which could be an indication that the model (\ref{zp},\ref{z})
does not reproduce DMFT even in the scaling limit \cite{contrast}.
From the model (\ref{zp},\ref{z}) one can also extract the asymptotic
behavior for large $x > \xi$
\begin{eqnarray}
  Z_x &\approx&\frac{ 0.8 \, U_c}{U_\text{right}-U_c} \exp\left[- \frac{x}{\xi}\right] \quad \text{for }
  x\to \infty \label{qpIns}
\end{eqnarray}
with
\begin{eqnarray}
  \xi &=& \sqrt{\frac{3}{22 \alpha}}
  \left(\frac{U_c}{U_\text{right}-U_c}\right)^{1/2} \approx
  0.09  \left(\frac{U_c}{U_\text{right}-U_c}\right)^{1/2}.\nonumber 
\end{eqnarray}

{\it Conclusions: } In this paper, we have studied how a metallic
state penetrates into a paramagnetic Mott insulator (or a bad metal).
Using a scaling analysis close to the quantum critical point we have
determined within dynamical mean field theory the critical exponents
and the asymptotic behavior of the quasiparticle weight close to and
far away from the interface.

The main physical mechanism governing the interface of a metal and a
Mott insulator is
the Kondo effect: the localized spins of the Mott insulator are
screened when they are brought into contact with the metal and become
therefore part of the metal. However, our numerical results show that
this mechanism is {\em not} very effective as can be seen from the
numerical prefactor in Eq.~(\ref{scaling}). Even for
$U_\text{right}=U_c$, the quasiparticle weight is only of size
$0.008/x^2$. There is no small parameter in the model which controls
this prefactor, which is reminiscent of another
small number characterizing the physics of Mott insulators: the
critical temperature describing the end-point of the first-order
Mott transition is much smaller than the Mott gap both within DMFT
\cite{BullaPRB2001} and in systems like V$_2$O$_3$ \cite{v2o3}.
Also the correlation length is extremely short: to obtain in
Eq.~(\ref{qpIns}) a correlation length of 10 lattice spacings, one has
to approach the critical point with a precision of $10^{-4}$. For all
practical purposes our results imply that the Mott insulator is de
facto impenetrable to the metal: Mott insulators are very good
insulators and the 'Kondo proximity effect' is inefficient. This is
consistent with our previous study of trapped fermionic atoms in an
optical lattice \cite{Helmes08}, where a metallic phase barely
penetrates into a coexisting Mott insulator.

The small quasiparticle weights at $T=0$ also imply that  very
small temperatures larger than the local Kondo temperature, $T_K
\propto Z$, efficiently quench the 'Kondo proximity
effect', see Fig.~\ref{fig1}.  Even more important is the effect of
magnetism which we have neglected in our study. The tiny local Kondo
temperatures in the Mott insulating phase will typically be much
smaller than the exchange couplings of the spins, wiping out the
Kondo effect. The magnetism of the Mott insulating phase will, in
contrast, penetrate easily into the metal \cite{Hofstetter08} via
Friedel oscillations of the magnetization.

For the future, it will be interesting to investigate with our methods
also models
which are not particle-hole symmetric where interface charges and long
range Coulomb interactions can lead to an electronic reconstruction of
the interface \cite{OkamotoMillis}.

\acknowledgements We thank R. Bulla, J. Kroha, Q. Liu, T. Micklitz, M.~Potthoff and M.~Vojta for
useful discussions. We acknowledge supercomputer support by the John von
Neumann institute for Computing (J\"ulich) and the Regional Computing
Center Cologne and financial support by the SFB 608 of the DFG.


\begin{thebibliography}{10}
\bibitem{shift} For good data collapse, we have shifted the $x$
  argument slightly, $x \to x-x_0$, where $x_0=-0.01$ for the DMFT+NRG
  results and $x_0=0.93$ for the Ginzburg-Landau style model for
  $U_\text{left}=2.59 D$ ($x_0$ depends on $U_\text{left}$).
\end{thebibliography}
\end{document}